\documentclass[trackchanges, twocolumn]{aastex701}

\usepackage{amsmath}
\begin{document}

\title{Suppression of inverse magnetic energy transfer in collisionless marginally magnetized plasmas}

\author[orcid=0000-0003-0961-7149,sname='Zhuo']{Zhuo Liu}
\affiliation{Plasma Science and Fusion Center, Massachusetts Institute of Technology, Cambridge, MA 02139, USA}
\email[show]{zhuol@mit.edu}  

\author[orcid=0000-0003-0709-7848
, sname='Zhou']{Muni Zhou} 
\affiliation{Department of Physics and Astronomy, Dartmouth College, Hanover, NH 03755, USA}
\email{}

\author[orcid=0000-0001-9755-6563,gname=Nuno,sname=Loureiro]{Nuno F. Loureiro}
\affiliation{Plasma Science and Fusion Center, Massachusetts Institute of Technology, Cambridge, MA 02139, USA}
\email{}

\begin{abstract}
We investigate the inverse cascade of magnetic energy in decaying, collisionless plasmas with moderate to high-$\beta$ values via first-principles numerical simulations and analytical theory. 
We find that pressure-anisotropy-driven instabilities, in particular the firehose instability, suppress reconnection-driven coalescence of magnetic structures (i.e., inverse transfer) by nullifying magnetic tension. 
This suppression leaves such structures elongated and confined to scales comparable to the Larmor radius of the particles. 
The presence of a magnetic guide field of sufficient strength, or a greater scale separation between the initial size of the magnetic structures and the Larmor radius, restores the system's ability to inverse transfer magnetic energy.
These results reveal that inverse energy transfer in collisionless plasmas is not guaranteed, but instead sensitively depends on magnetization. In the astrophysical context, this identifies a kinetic mechanism by which Weibel-generated seed fields may fail to merge consistently, potentially limiting their role in cosmic magnetogenesis.
\end{abstract}

\section{Introduction} 

The origin and evolution of cosmic magnetic fields remain fundamental unsolved problems in astrophysics and plasma physics. 
Observations of coherent, large-scale magnetic fields in galaxy clusters and the intergalactic medium (IGM) imply that weak, primordial ``seed'' fields must have been generated initially and subsequently amplified by turbulent dynamo processes to reach their observed strengths (e.g., \citet{brandenburg2005astrophysical,kulsrud2008origin}). 
However, the mechanisms responsible for the generation and dynamics of these seed fields, particularly in collisionless, high-beta plasmas typical of many astrophysical environments, remain poorly understood.

A promising mechanism for generating such ``seed'' magnetic fields is the Weibel instability, a kinetic instability driven by anisotropic velocity distributions in an unmagnetized plasma that produces magnetic filaments on microscopic scales \citep{weibel1959spontaneously,medvedev1999generation,schlickeiser2003cosmological}. 
The Weibel instability is relevant in weakly collisional astrophysical plasmas, such as those in the intracluster medium (ICM) and IGM, which are largely unmagnetized at early times \citep{ryu1998cosmic,brandenburg2023galactic}. 
Theoretical and numerical studies have shown that the Weibel instability converts a fraction of the free energy due to velocity-space anisotropy into filamentary magnetic fields and saturates once the plasma becomes magnetized (e.g., when the particle bounce/cyclotron frequency becomes comparable to the linear growth rate)~\citep{weibel1959spontaneously, Fried1959,Califano1998, Grassi2017, Bresci2022}. 
However, for these small-scale fields to contribute meaningfully to large-scale magnetogenesis, they likely need to undergo an inverse cascade, merging into progressively larger-scale structures before further amplification by turbulent dynamo action \citep{brandenburg2001inverse,rincon2016turbulent,vazza2018resolved,zhou2023magnetogenesis}.

The inverse transfer of magnetic energy in a decaying turbulent plasma has been widely studied within the resistive magnetohydrodynamics (MHD) framework, where it is generally attributed to magnetic-reconnection-mediated merging of magnetic structures \citep{zrake2014inverse, brandenburg2015nonhelical,zhou2019magnetic, zhou2020multi, hosking2021reconnection}. 
However, this picture does not necessarily apply to the collisionless, unmagnetized plasma conditions where the Weibel instability operates. 
The Weibel instability generates fields at scales comparable to the Larmor radius of thermal particles, making a purely MHD treatment inadequate to describe their subsequent evolution.
In addition, kinetic effects such as the firehose and mirror instabilities can play a crucial role in regulating the evolution of magnetic structures by modifying the effective viscosity of the plasma \citep{kunz2014firehose, melville2016pressure, st2018fluctuation}. 
A fully kinetic approach is thus required to determine whether these seed fields, undergoing an inverse cascade, can continuously coalesce and reach progressively larger scales, or whether instead the inverse coalescence is arrested by kinetic effects or instabilities, resulting in the confinement of the magnetic fields to microscopic scales.

This Letter directly investigates a crucial missing stage in the magnetogenesis process: the coalescence and inverse transfer of magnetic fields within the Weibel-seed context before they enter the turbulent dynamo regime. 
While previous studies have focused on the generation and saturation of these fields \citep{brandenburg2015nonhelical,zhou2022spontaneous, zhou2023magnetogenesis} or their subsequent amplification by dynamo processes \citep{brandenburg2005astrophysical, st2018fluctuation,st2020fluctuation,sironi2023generation, zhou2023magnetogenesis}, the intermediate phase, i.e., how these small-scale fields merge into larger structures, has remained largely unexplored due to computational limitations. 
To address this gap, we conduct fully kinetic simulations that initialize a weakly magnetized, moderate to high-beta plasma in a state directly relevant to the saturation phase of the Weibel instability, and follow its evolution. 
Understanding this transition is essential for a comprehensive theory of magnetogenesis, with implications for the evolution of magnetic fields in galaxy clusters, the intergalactic medium, and other weakly collisional astrophysical systems.

\section{Numerical Setup} 
We perform first-principles particle-in-cell (PIC) simulations using the fully relativistic code \textsc{Osiris}~\citep{Fonseca2002, Hemker2015}.
To reduce computational cost, the simulations employ an electron-positron plasma ($s \in {e,p}$), and are only 2.5-dimensional, i.e., all fields and particles' positions depend only on the in-plane coordinates $(x, y)$, while the three components of particle velocity and electromagnetic fields are retained. 
Periodic boundary conditions are applied in both directions. 
The computational domain has size $L_x \times L_y$ with $L_x = L_y = L$, discretized with $1024 \times 1024$ grid cells.
The initial configuration consists of a $2k_0 \times 2k_0$ static array of magnetic islands with alternating polarities (and, therefore, zero net flux in the simulation box). 
The in-plane magnetic field is given by
$B_x = -B_0\cos(2\pi k_0x/L)\sin(2\pi k_0y/L)$ and $B_y = B_0 \sin(2\pi k_0x/L)\cos(2\pi k_0y/L)$,
where $B_0$ is the maximum initial in-plane magnetic field amplitude and $k_0$ is the wavenumber of the island array.
We initialize the particles with Maxwellian velocity distributions of temperature $T_0$.
The particle density consists of a uniform background $n_0$ plus a spatial variation
$n_s = (B_0^2 / 2 T_0) \cos^2(2\pi k_0 x / L) \cos^2(2\pi k_0 y / L)$,
chosen to satisfy the initial force balance $\mathbf{J} \times \mathbf{B}/c = \nabla p$. 
The out-of-plane current density is
$J_z = e n_0 (u_{d,p0} - u_{d,e0}) = 2 e n_0 u_{d,p0} = (2\pi k_0 / L) \cos(2\pi k_0 x / L) \cos(2\pi k_0 y / L)$, consistent with Amp\'ere's law. 

All quantities are normalized in PIC units: time to $\omega_p^{-1}$, length to $d_0 = c / \omega_p$, density to $n_0$, temperature to $m c^2$, and magnetic field to $c m \omega_p / e$, where $\omega_p= \sqrt{4\pi n_0 e^2/m}$ is the plasma frequency corresponding to $n_0$. 
We set $B_0 = 0.5$, yielding a magnetization parameter $\sigma = B_0^2 / (4\pi n_0 m c^2) = 0.25$, and $T_0 = 0.25$ so that particle motions remain subrelativistic.
The initial plasma beta is $\beta_0 = \langle 8\pi p \rangle / \langle B^2 \rangle \approx 10.8$ (the operator $\langle . \rangle$ denotes volume average).
The Larmor radius, $\rho_0 = \langle v_{\mathrm{th}} \rangle / \langle \Omega \rangle$, is approximately $2.1$, where $v_{\mathrm{th}}$ is the thermal velocity and $\langle \Omega \rangle$ is computed using the root mean squared of the total magnetic field. 
We set $L=80\pi$ and $k_0 = 16$ (note that $k_0$ is a dimensionless parameter), so that the initial island scale is approximately $L/ (2\pi k_0) = 2.5$, placing the plasma in a marginally magnetized regime ($R_0 \approx 0.8\rho_0$).

\section{Inverse cascade without a guide magnetic field}

\begin{figure}[!htbp]
\includegraphics[width=0.49\textwidth]{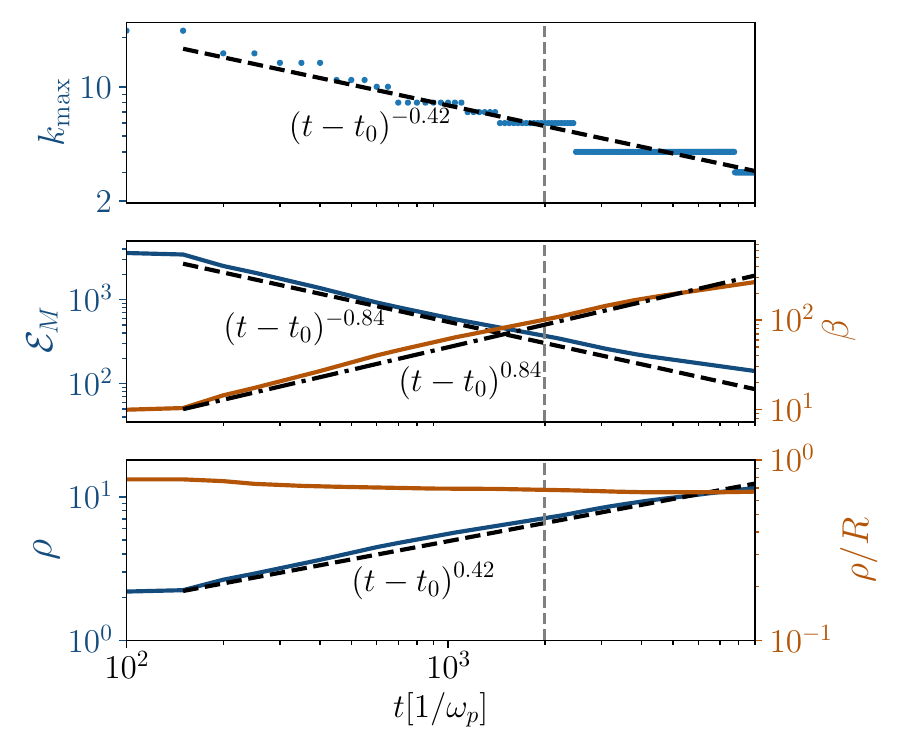}
\caption{\label{fig:1.power laws}  Time evolution of the energy-containing wavenumber $k_{\rm max}$ (top), magnetic energy $\mathcal{E}_M$ and plasma beta $\beta$ (middle), and the Larmor radius $\rho$ as well as the ratio $\rho/R$ between the Larmor radius and the average magnetic structure size (bottom). The vertical dashed line marks the approximate onset of significant pressure anisotropy effects. }
\end{figure}

The system described above is an unstable equilibrium that will evolve and gradually decay. 
When same-polarity current filaments approach one another, they merge through magnetic reconnection of their in-plane magnetic fields.
Each generation of mergers reduces the number of filaments by half while increasing the characteristic size of the resulting magnetic structures.
As this hierarchical process repeats, progressively larger-scale magnetic structures form, driving an inverse transfer of magnetic energy.
This cascade is self-similar, leading to power-law temporal scalings of key system quantities proportional to $(t - t_0)^{\alpha}$, where $\alpha$ is a scaling exponent and $t_0$ is the time at which the first generation of mergers starts to occur.
A theoretical framework for describing this process (in the context of MHD) was first introduced by \citet{zhou2019magnetic}; see also \citet{zhou2020multi, zhou2021statistical} for further details.

Fig.~\ref{fig:1.power laws} presents the temporal evolution of several key quantities. 
The top panel shows the wavenumber $k_{\rm max}$ at which the magnetic energy spectrum peaks. 
Its monotonic decrease reflects the ongoing inverse transfer of magnetic energy driven by magnetic island mergers, following a power-law scaling of approximately $k_{\rm max} \propto (t - t_0)^{-0.42}$. 
Previous studies have reported a $t^{-0.5}$ scaling for this quantity in reduced MHD~\citep{zhou2019magnetic, zhou2020multi, zhou2021statistical}, full MHD~\citep{bhat2021inverse}, and kinetic simulations in the reduced MHD regime~\citep{liu2025electron}. 
We attribute the slightly shallower scaling observed here to the (deliberately) limited scale separation between the reconnecting magnetic structures and the particles' Larmor radius (see Appendix~\ref{app:sec:3} for validation).

In a self-similar merger cascade, the total magnetic energy is expected to scale with the square of the characteristic wavenumber~\citep{zhou2019magnetic, hosking2021reconnection}. 
Thus, for $k_{\rm max} \propto (t - t_0)^{-0.42}$, we predict $\mathcal{E}_M \propto (t - t_0)^{-0.84}$, consistent with the evolution shown in Fig.~\ref{fig:1.power laws} (middle panel). 
As the magnetic energy decays, the plasma beta correspondingly increases, following $\beta \propto 1/\mathcal{E}_M \propto (t - t_0)^{0.84}$. 
This behavior reflects the high-$\beta$ regime of the system, where the thermal pressure remains approximately constant due to the limited amount of magnetic energy available for plasma heating.
Based on these results, the temporal evolution of the volume-averaged in-plane magnetic field at $t>t_0$ can be approximated as (assuming $B_{x,y}(t=t_0) \approx B_0/\sqrt{2}$)
\begin{equation}
\label{eq:Bfield}
\langle B_{x,y} \rangle (t) \approx \frac{B_0}{\sqrt{2}} \left( 1+  \left(\frac{t-t_0 }{\tau_0} \right)\right)^{-\alpha},
\end{equation}
where $\alpha \approx 0.42$ is the measured scaling exponent and $\tau_0$ is the reconnection time of the first generation of mergers~\citep{zhou2020multi}.
When $(t-t_0)/\tau_0 \gg 1$, Eq.~\eqref{eq:Bfield} shows that $\langle B_{x,y} \rangle (t) \propto (t-t_0)^{-\alpha}$.
The bottom panel of Fig.~\ref{fig:1.power laws} shows that the averaged particle Larmor radius $\rho$ (blue) increases in time as the magnetic field weakens, following approximately $(t - t_0)^{0.42}$.
This scaling matches that of the characteristic magnetic structure size, $R \propto 1/k_{\rm max}$, as expected, since $\rho \propto 1/B$ in the absence of significant particle heating.
Taking $\rho/R$ (or, equivalently, $k_{\rm max} \rho$) as a measure of the system’s magnetization, the near constancy of this ratio over time indicates that the plasma stays marginally magnetized throughout the evolution.
This is confirmed by the orange curve in the bottom panel of Fig.~\ref{fig:1.power laws}, which shows $\rho / R \approx 0.8$ for the duration of the simulation.

In magnetized high-$\beta$ collisionless plasmas, the perpendicular and parallel pressures evolve according to the conservation of two single-particle adiabatic invariants: the magnetic moment, $\mu = m |\mathbf{w}_{\perp}|^2 / (2B)$, which governs perpendicular dynamics, and the longitudinal invariant, $\mathcal{J} = \oint m \mathbf{w}_{\parallel} \cdot d\mathbf{x}$, which constrains parallel motion (here, $\mathbf{w}_{\perp}$ and $\mathbf{w}_{\parallel}$ are the thermal velocity perpendicular and parallel to the local magnetic field, respectively, of any given particle).
During magnetic island mergers, a local decrease in $B$ reduces $w_{\perp}$, leading to adiabatic cooling of $T_\perp$ and the generation of negative anisotropy, $\Delta \equiv T_\perp/T_\parallel - 1 < 0$. 
At the same time (but not necessarily at the same locations), the merging process of the islands creates elongated magnetic structures so that the path length of particles along the field lines increases, thereby decreasing $w_{\|}$ and locally yielding $\Delta >0$ within the merging structures.
We note that the system contains magnetic null points (reconnection X-points),
where single-particle adiabatic invariants are not conserved. 
Moreover, the average Larmor radius of the particles is only marginally smaller than the characteristic magnetic length scale, so the adiabatic invariants should only be approximately conserved.

As the plasma beta increases during the inverse cascade, the thresholds for anisotropy–driven instabilities become easier to exceed because they scale inversely with $\beta_\parallel$: $\Delta < -2/\beta_\parallel$ for firehose and $\Delta > 1/\beta_\parallel$ for mirror. Even modest anisotropies can therefore destabilize the plasma at late times. Once excited, these microinstabilities scatter particles and reduce the effective magnetic tension \citep{schoeffler2013role,kunz2014firehose,rincon2015non}, constraining the system to drift in the stable region of the $(\Delta,\beta_\parallel)$ plane and thereby slowing reconnection–driven mergers and suppressing the inverse cascade.


\begin{figure}[!htbp]
\includegraphics[width=0.49\textwidth]{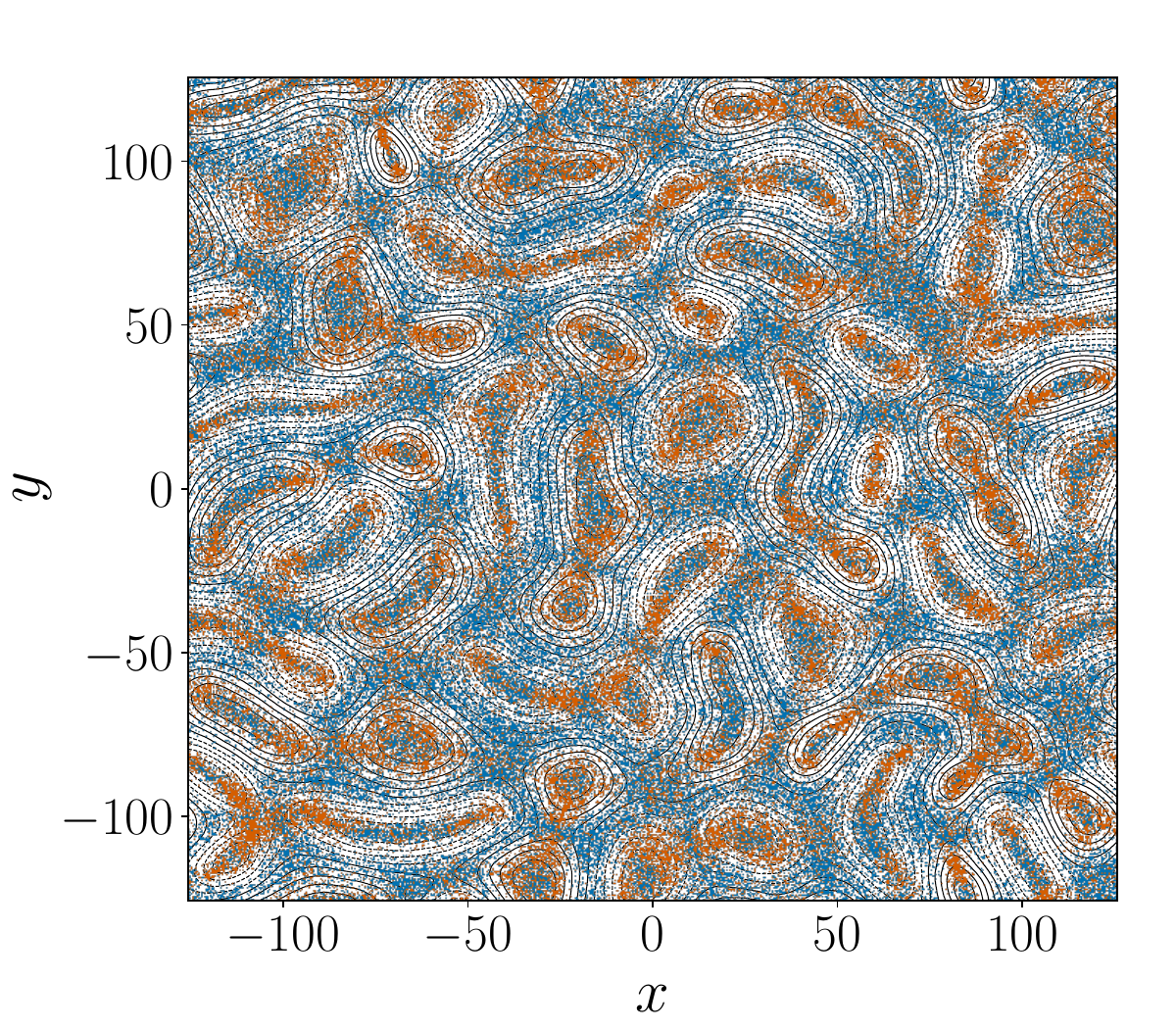}
\caption{\label{fig:2.unstable} Contours of magnetic flux at $\omega_{p}t = 2500$. Blue and orange identify the locations unstable to the firehose or mirror instabilities, respectively.}
\end{figure}

Consistent with this expectation, at later times ($\omega_p t \gtrsim 2500$, marked by the dashed line in Figure~\ref{fig:1.power laws}), the system deviates from its earlier power-law behavior and the inverse cascade slows (see Fig.~\ref{fig:1.power laws}). 
Fig.~\ref{fig:2.unstable} further shows magnetic-flux contours at $\omega_p t = 2500$: islands are elongated, evidencing reduced magnetic tension and slower contraction. 
Mirror unstable regions (orange) concentrate around elongated merging islands where extended bounce paths favor $\Delta > 0$, while firehose unstable regions (blue) appear near reconnection sites and at the edges of merged islands where $B$ is weak and $\Delta < 0$. 
Similar island elongation and reconnection suppression have been reported in systems initialized with multiple Harris sheets~\citep{schoeffler2013role}.

Additional diagnostics at $\omega_p t = 10000$ (first column of Fig.~\ref{fig:3.contour}) show smooth magnetic-flux contours interspersed with out-of-plane current concentrations at scales smaller than the islands (panel (a)), consistent with firehose/mirror activity injecting energy near the Larmor scale at later times.
Correspondingly, the magnetic-energy spectra at progressively later times (panel (d)) develop plateaus to the right of the spectral peak, indicating small-scale energy injected by these instabilities; the injection scale shifts to larger scales as the islands grow because $\rho/R \approx \mathrm{const.}$
Pressure anisotropy evolution is shown in panel (g) through distributions of $\Delta$ versus $\beta_\parallel$.
At later times, more regions cross the mirror and firehose instability thresholds. 

However, as expected, the bulk of the distribution remains clustered near the stable region ($\Delta = 0$). 
This implies that, although localized regions may intermittently become unstable, a significant portion of the domain stays within the stable region between the firehose and mirror thresholds. This allows island merging and inverse energy transfer to proceed, albeit less efficiently than in the absence of pressure anisotropy effects.
We note that in the presence of a guide magnetic field, the behavior is expected to be markedly different. 
In that case, a decrease in the (in-plane) magnetic field due to mergers is expected to drive a system-averaged negative pressure anisotropy. 
We investigate this in the following section.

\section{Inverse cascade with a guide magnetic field}
In this section, we extend our study of decaying turbulence to plasmas threaded by a uniform out-of-plane (guide) magnetic field, $\mathbf{B}_G = B_G \hat{z}$. 
With a finite guide field, the system contains no magnetic null points. 
As coalescence proceeds and the in-plane magnetic field component weakens, the local field becomes increasingly aligned with the $z$–axis.
Conservation of the magnetic moment, $\mu$, still leads to cooling of perpendicular motions as in-plane $B$ weakens. 
At the same locations, conservation of energy (assuming kinetic energy injected by reconnection is small compared to the thermal energy of the particles in high beta plasma) converts the loss of magnetic-mirror potential energy ($\mu B$) into parallel kinetic energy, thereby increasing both $w_{\parallel}$ and $T_{\parallel}$.
This effect is expected to apply across the entire domain (due to the absence of magnetic null points) and, therefore, generate a \textit{system-averaged} negative pressure anisotropy ($\Delta < 0$) throughout each merger.

\begin{figure*}[!htbp]
\includegraphics[width=1.0\textwidth]{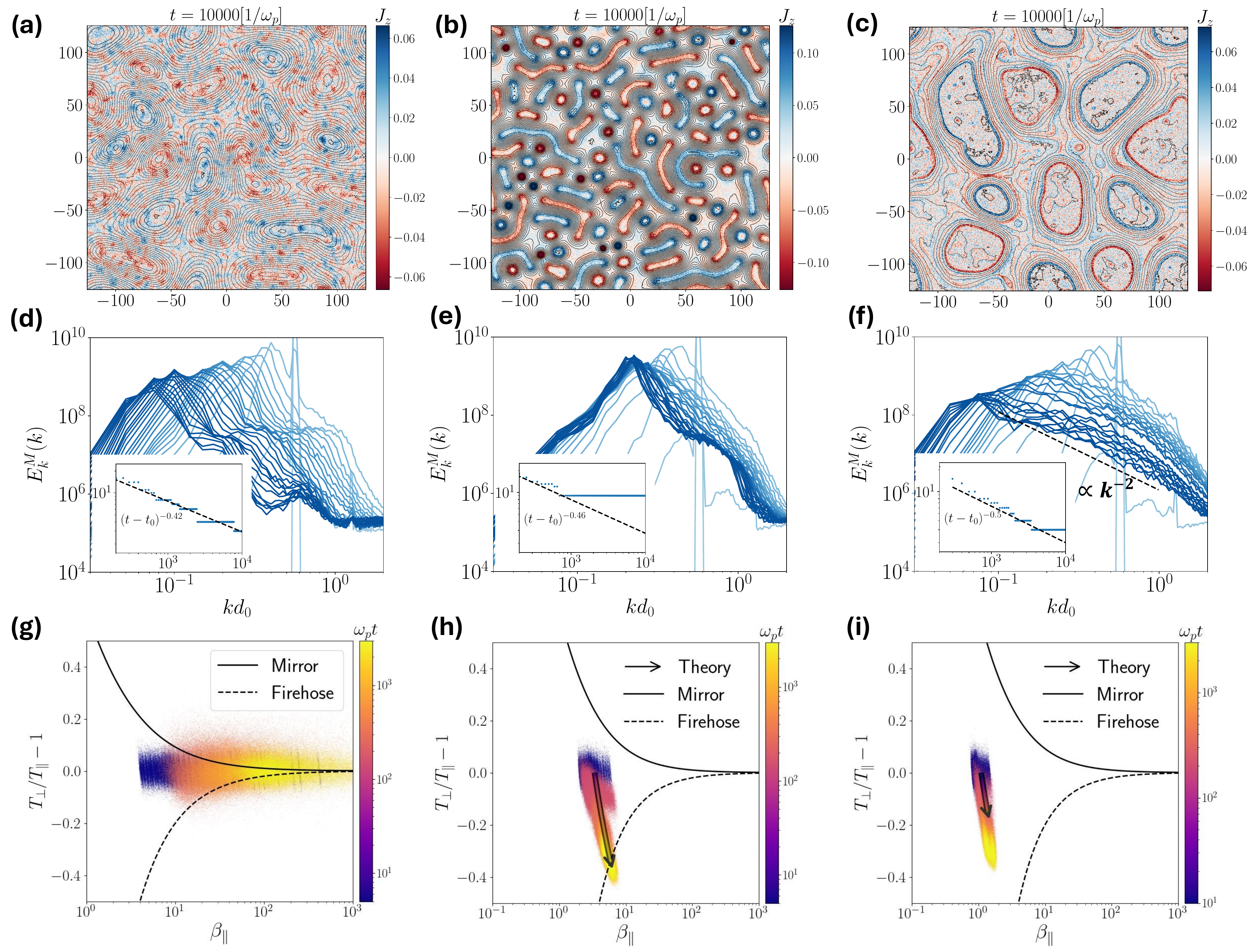}
\caption{\label{fig:3.contour} Top row (a)–(c): Out-of-plane current density (color scale) and magnetic flux contours (curves) at $\omega_p t = 10000$ for simulations with guide field strengths $\hat{B}_G = 0$, $1.0$, and $2.0$, respectively.
Middle row (d)–(f): Magnetic energy spectra at successive times for each corresponding case. Insets show the temporal evolution of the wavenumber $k$ at which the magnetic energy spectrum peaks.
Bottom row (g)–(i) \footnote{For clarity of visualization, the time evolution of the scatter plots is truncated at $3000\omega_p^{-1}$, after which no significant changes are observed.}: Scatter plots of pressure anisotropy $T_{\perp}/T_{\|} - 1$ versus parallel plasma beta $\beta_{\|}$ at progressively later times, indicated by color. In panels (h) and (i) \footnote{In panel (i), the theoretical trajectory is shorter compared to the observation. This is because the CGL-MHD equations with the heat fluxes neglected do not hold at the initial transient stage of the simulation; see Appendix~\ref{app:sec:1} for evidence.}, arrows denote theoretical predictions for the anisotropy evolution trajectories (see Appendix~\ref{app:sec:1} for details).} 
\end{figure*}

In this scenario, once the firehose threshold $\Delta = -2/\beta_{\parallel}$ is reached, the state of the system in the $(\Delta,\beta_{\parallel})$ plane will cluster around the firehose instability boundary with system-averaged negative pressure anisotropy, leaving no nearby stable region into which the mergers can proceed directly.
Furthermore, because the plasma is only marginally magnetized, the driving operates at scales comparable to those of the most unstable firehose modes ($k\rho \sim 1$). 
Once the threshold $\Delta \leq -2/\beta_\parallel$ is locally met, the effective magnetic tension vanishes, $v_{A,\rm eff}=v_A\sqrt{1+\beta_{\|}\Delta/2}\to 0$, stopping coalescence (because $k\rho\sim kR\sim1$). 
In other words, the system effectively stalls near marginal stability, directly resulting in a ``frozen'' scenario.

To test these predictions, we repeat the simulations with a uniform guide field at strengths $\hat{B}_G=1.0$ and $\hat{B}_G=2.0$ (with $\hat{B}_G\equiv B_G/B_0$), keeping all other parameters fixed. 
The corresponding initial plasma betas are $\beta_0 \approx 3$ and $\beta_0 \approx 1$, and the initial Larmor radii are $\rho_0 \approx 1.2$ and $\rho_0 \approx 0.7$, respectively. 
The top row of Fig.~\ref{fig:3.contour} shows magnetic flux contours (curves) and out-of-plane current density (color) at $\omega_p t = 10000$ for simulations with $\hat{B}_G = 0$, $1.0$, and $2.0$ (left to right). In both the $\hat{B}_G = 0$ and $\hat{B}_G = 2.0$ cases, magnetic structures grow to the box scale, consistent with sustained inverse energy transfer. 
In contrast, the $\hat{B}_G = 1.0$ run develops smaller, highly elongated, ``worm''-like structures, providing direct evidence for the nullification of magnetic tension and the ``frozen’’ scenario described above, which we attribute to the triggering of the firehose instability.

Spectral diagnostics corroborate this picture. 
In panel (e) of Fig.~\ref{fig:3.contour}, the magnetic-energy spectrum for $\hat{B}_G = 1.0$ ceases to shift to lower $k$ once the peak reaches $k d_0 \approx 0.2$, while a plateau forms near $k d_0 \approx 0.5$, signaling power injection at Larmor scales by firehose fluctuations. 
By contrast, the $\hat{B}_G=2.0$ spectrum continues to evolve. 
Free from disruption by pressure-anisotropy-driven instabilities, it develops a $k^{-2}$ scaling, consistent with the formation of sharp, localized current sheets as reported previously in 2D studies~\citep{zhou2019magnetic}.

The bottom row of Fig.~\ref{fig:3.contour} clarifies why the case with $\hat{B}_G = 1.0$ becomes firehose unstable while $\hat{B}_G = 2.0$ does not. 
These panels show scatter plots of $\Delta$ versus $\beta_{\|}$ at progressively later times up to $\omega_p t = 3000$. 
As the mergers proceed, $\beta_{\|}$ steadily increases and the anisotropy becomes increasingly negative, such that the bulk distribution drifts toward the lower right of these plots (panels h–i).

Whether the system ultimately crosses the firehose boundary depends on the initial plasma beta $\beta_0$ and the strength of the guide field $B_G$.
To make a more quantitative prediction, we estimate the volume-averaged magnetic-field strength as $\langle B \rangle (t)=\sqrt{\langle B_{x,y}\rangle (t)^2+B_G^2}$, with $B_{x,y}(t)$ given by Eq.~\eqref{eq:Bfield}.
We adopt the Chew-Goldberger-Low MHD (CGL-MHD) equations (with heat fluxes neglected)~\citep{chew1956boltzmann} to describe the system-averaged evolution: $\langle p_\perp \rangle /\langle n \rangle \langle B\rangle \approx \mathrm{const.}$ and $\langle p_\parallel \rangle\langle B\rangle^2/\langle n\rangle^3 \approx \mathrm{const}$.
We find that avoiding the firehose condition requires a sufficiently low $\beta_0$ and a sufficiently strong guide field. 
This requirement can be formalized through the expression (see Appendix~\ref{app:sec:1} for the detailed derivation)
\begin{equation}
\label{eq:G}
G(\hat{B}_G,\beta_0) \equiv \left(\frac{2\hat{B}_G^2}{2\hat{B}_G^2+1}\right)^{3/2}
+ \frac{2}{\beta_0}\left(\frac{2\hat{B}_G^2}{2\hat{B}_G^2+1}\right)^2 - 1.
\end{equation} 
If $G(\hat{B}_G,\beta_0) \leq 0$, the firehose condition will inevitably be met.
Since $G(\hat{B}_G,\beta_0)$ increases monotonically with $\hat{B}_G$, there exists a critical guide field above which the cascade can proceed unimpeded for a fixed $\beta_0$.
Moreover, the condition $G(\hat{B}_G,\beta_0) > 0$ is more difficult to reach with a larger $\beta_0$, indicating that a system with a larger initial plasma beta is more likely to trigger the firehose instability.
In our simulations $G(\hat{B}_G,\beta_0) < 0$ for $\hat{B}_G = 1.0$, consistent with the ``frozen'' evolution seen in Fig.~\ref{fig:3.contour} (b,e,h), while $G(\hat{B}_G,\beta_0) > 0$ for $\hat{B}_G = 2.0$, consistent with the uninterrupted cascade in Fig.~\ref{fig:3.contour} (c,f,i). 

The analytic evolutionary trajectories in the $(\beta_{\parallel}, \Delta p/p)$ plane can also be derived from the CGL-MHD equations (see Appendix~\ref{app:sec:1}) and are shown as black arrows in panels (h) and (i), with $t_0$ and $\tau_0$ obtained by fitting the decay of the in-plane magnetic field. 
The close agreement between these predictions and the simulation results confirms that the systems' progression toward the firehose threshold is well captured by our analytical model. 
Because the system becomes ``frozen'' once the firehose condition is reached, the extent of inverse transfer in firehose-unstable cases remains minimal, with the final characteristic wavenumber $k^*$ staying close to its initial value $k_0$ for a large range of $\beta_0$ (see Appendix~\ref{app:sec:1} for further evidence).

Finally, we note that we expect the ``frozen'' state to occur only in marginally magnetized plasmas with $\rho \sim R$, as argued above. 
When there is a clear scale separation ($R \gg \rho$), the inverse cascade should not stall completely: the firehose instability acts and saturates at $k\rho \sim 1$, where pitch–angle scattering relaxes the anisotropy to marginal stability on the Larmor scale, while the magnetic tension on the reconnection scale ($\sim R$) remains finite. 
As a result, reconnection continues to drive island coalescence, and the system evolves along the firehose marginal–stability boundary rather than becoming arrested. 
We confirm this hypothesis in Appendix~\ref{app:sec:2}.

\section{Conclusions} 
Using fully kinetic simulations of decaying, marginally magnetized, high-$\beta$ pair plasmas, we have shown that the inverse transfer of magnetic energy via the coalescence of magnetic structures can be strongly impeded by pressure-anisotropy-driven microinstabilities, most notably the firehose. 
When the effective magnetic tension is reduced near the firehose boundary, island contraction and merger stall, resulting in highly elongated structures and an arrest of the inverse cascade.
This effect becomes more prominent when the plasma beta is large.
These results identify a kinetic mechanism by which Weibel–generated seed fields may fail to grow coherently by coalescence in collisionless, high-$\beta$ environments.

Our conclusions fit naturally into the broader kinetic picture of cosmic magnetogenesis \citep[e.g.,][]{zhou2023magnetogenesis}, which may involve seed generation by the Weibel instability, inverse transfer via reconnection-driven coalescence, and turbulent dynamo amplification. 
The present work shows that the second stage (inverse transfer) can be effectively suppressed in the absence of a sufficiently strong guide or mean field.
For Weibel-seeded fields with initial $\beta_0^{-1} \sim (L/d_e)^{-1/2} M^{1/4}$ \citep{zhou2023magnetogenesis}, typical intracluster medium (ICM) parameters ($M \sim 0.1$, $L/d_e \sim 10^{15}$) yield $\beta_0 \sim 10^8$, implying that significant inverse transfer is unlikely. 
In such cases, the Weibel-generated fields remain confined to their microscopic coherence scale, which may in turn hinder further turbulent dynamo amplification, unless the turbulent dynamo can pick up seed fields directly at electron-kinetic scales.

Several caveats should be noted. 
First, this study is restricted to two spacial dimensions, which can permit long-lived metastable equilibria whose existence in fully turbulent, three-dimensional systems is questionable.
Three-dimensional kinetic simulations are required to assess whether additional instability channels can dynamically regulate pressure anisotropy, reduce the residence time in firehose-unstable states, and alter the long-term evolution of electron-scale seed fields.
Second, we employ a pair plasma for computational efficiency. In an electron–ion plasma, both species can excite Weibel instabilities and saturate at their respective kinetic scales. The influence of magnetized electrons on ion-Weibel instability remains poorly understood, as does the role of electron-driven pressure anisotropy and associated instabilities in modifying the coalescence of seed fields at ion-Larmor scales. Addressing these open questions will be essential for connecting kinetic-scale physics with the dynamics of cosmic magnetogenesis.

\begin{acknowledgments}
This work was supported by the Mathworks Fellowship (ZL), the U.S. Department of Energy under contract number DE-FG02–91ER-54109 (ZL and NFL), and by NSF Grant No.~2512037 (MZ).
The authors extend their gratitude to Dmitri Uzdensky and Greg Werner for valuable discussions and insights.
This research used resources of the MIT-PSFC partition of the Engaging cluster at the MGHPCC facility, funded by DOE award No. DE-FG02-91-ER54109, and the National Energy Research Scientific Computing Center, a DOE Office of Science User Facility supported by the Office of Science of the U.S. Department of Energy under Contract No. DE-AC02-05CH11231 using NERSC award FES-ERCAP0020063.

\end{acknowledgments}

\appendix

\section{Well magnetized scenario without a guide field\label{app:sec:3}}

\begin{figure}[!htbp]
\centering
\includegraphics[width=0.6\textwidth]{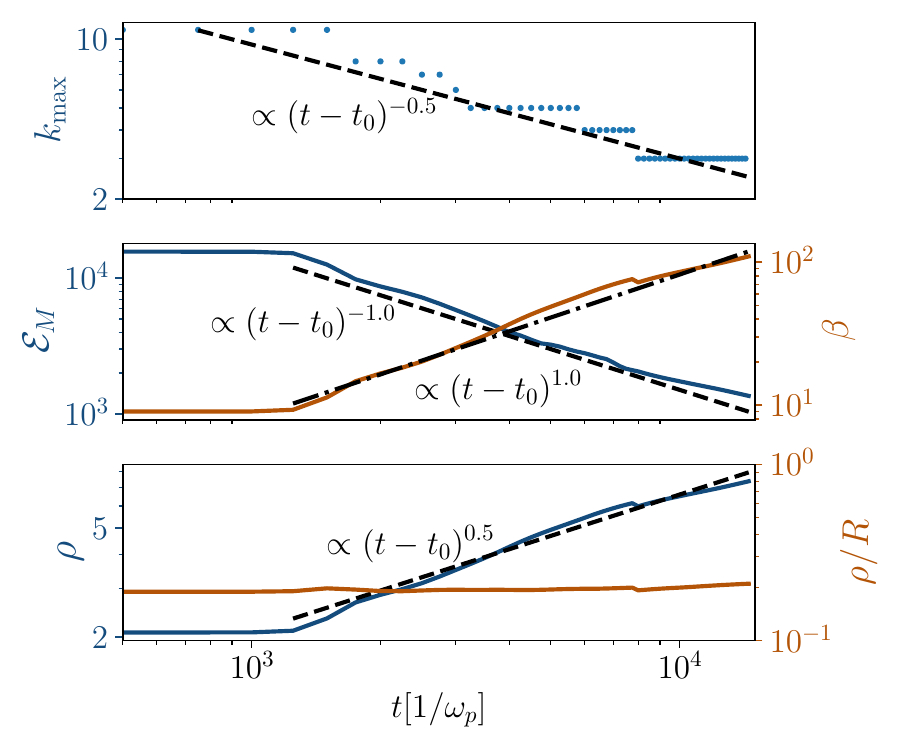}
\caption{\label{app:fig:5.trace_mhd} Time evolution of energy containing scale $k_{\rm max}$ (top panel), magnetic energy $\mathcal{E}_M$ and plasma beta $\beta$ (middle panel), Larmor radius $\rho$, and ratio between Larmor radius and averaged magnetic structure size $\rho/R$ (bottom panel) from the well magnetized simulations without a guide magnetic field. }
\end{figure} 

To assess whether the deviations of the power-law scalings reported in the main text from those in previous MHD simulations originate from the limited scale separation between the particle Larmor radius and the size of magnetic structures, we repeat the simulation from Section~2 with twice the box size ($L_x = L_y = 160\pi$) and four times the initial island size ($k_0 = 8$), while keeping the initial Larmor radius fixed. 
The simulation domain is described with $2048 \times 2048$ grid cells.
This configuration increases the scale separation and moves the system closer to the MHD regime, allowing for a more direct comparison with MHD expectations.

Fig.~\ref{app:fig:5.trace_mhd} shows the time evolution of the key global quantities in this simulation. 
The energy-containing wavenumber $k_{\rm max}(t)$ follows a power-law decay with an exponent of $-0.5$, while the magnetic energy $\mathcal{E}_M(t)$ decays as $t^{-1}$. 
These scaling behaviors align well with predictions from MHD simulations of magnetic island mergers \citep[e.g.,][]{zhou2020multi, hosking2021reconnection}. 
The bottom panel shows that the ratio between the Larmor radius and the average magnetic structure size ($\rho/R$) remains approximately constant at $0.2$, indicating sustained scale separation throughout the evolution.

\section{Evolution of pressure anisotropy and parallel beta\label{app:sec:1}}

In this appendix we derive an analytical model for the evolution of the system-averaged pressure anisotropy and plasma beta in our system.
We begin with the assumption that the evolution of the system follows the CGL-MHD equations with heat fluxes neglected:
\begin{align}
    \frac{d}{dt} \left( \frac{ p_{\perp} }{ n  B} \right) = 0,  \\
\frac{d}{dt} \left( \frac{ p_{\parallel} B^2}{ n^3} \right) = 0;
\end{align} 
and we further assume that these conservation laws approximately hold in a volume averaged sense:
\begin{align}
    \frac{d}{dt} \left( \frac{\langle p_{\perp}\rangle}{\langle n \rangle \langle B\rangle} \right) \approx 0,  \\
\frac{d}{dt} \left( \frac{\langle p_{\parallel} \rangle \langle B\rangle^2}{\langle n\rangle^3} \right) \approx 0.
\end{align} 

Then we assume the density remains relatively constant in time and find
\begin{align}
\langle T_{\|} \rangle (t) \approx \frac{\langle B(t_0) \rangle^2}{\langle B(t) \rangle^2} T_0, \\
    \langle T_{\perp}\rangle  (t) \approx \frac{\langle B(t) \rangle}{\langle B(t_0) \rangle} T_0,
\end{align}
where we have assumed $\langle T_{\|,\perp} \rangle (t_0) \approx T_0$.

By plugging in Eq.~\eqref{eq:Bfield}, the temporal evolution of pressure anisotropy is expected to follow
\begin{equation}
\label{app:eq:delta}
   \langle \Delta \rangle (t) \approx \frac{\langle T_{\perp}\rangle (t)}{\langle T_{\|}\rangle (t)}  - 1  \approx \left(\frac{\langle B(t) \rangle}{\langle B (t_0)\rangle}\right)^3 - 1 \approx  \left(\frac{2 \hat{B}_G^2+\left( 1+  \left(\frac{t-t_0 }{\tau_0} \right)\right)^{-2\alpha}}{2\hat{B}_G^2+1}\right)^{3/2} -1,
\end{equation}
and the temporal evolution of parallel plasma beta can be estimated as
\begin{equation}
\label{app:eq:beta}
    \langle \beta_{\parallel} \rangle (t) \approx \frac{8\pi n_0 \langle T_{\|} \rangle (t)}{\langle B(t) \rangle^2} \approx \beta_0\left(\frac{ \langle B(t_0) \rangle}{\langle B(t) \rangle}\right)^4\approx \beta_0 \left(\frac{2\hat{B}_G^2+1}{2 \hat{B}_G^2+\left( 1+  \left(\frac{t-t_0 }{\tau_0} \right)\right)^{-2\alpha}}\right)^2,
\end{equation}
where we have assumed that $\langle \beta_{\|} \rangle (t_0) \approx \beta_0$.
Eqs.~\eqref{app:eq:delta} and~(\ref{app:eq:beta}) together define a trajectory in time as a function of input parameters $\beta_0$, $t_0$, and $\hat{B}_G$.
Then it is not hard to find that the firehose condition $\langle \Delta \rangle \leq - 2/\langle \beta_{\|} \rangle$ will eventually be met ($t\rightarrow \infty$) if $G(\hat{B}_G,\beta_0) \leq 0$, where
\begin{equation}
G(\hat{B}_G,\beta_0) =  \left(\frac{2\hat{B}_G^2}{2\hat{B}_G^2+1}\right)^{3/2} + \frac{2}{ \beta_0}\left(\frac{2\hat{B}_G^2}{2\hat{B}_G^2+1}\right)^2 - 1.
\end{equation}

\begin{figure}[!htbp]
\centering
\includegraphics[width=0.9\textwidth]{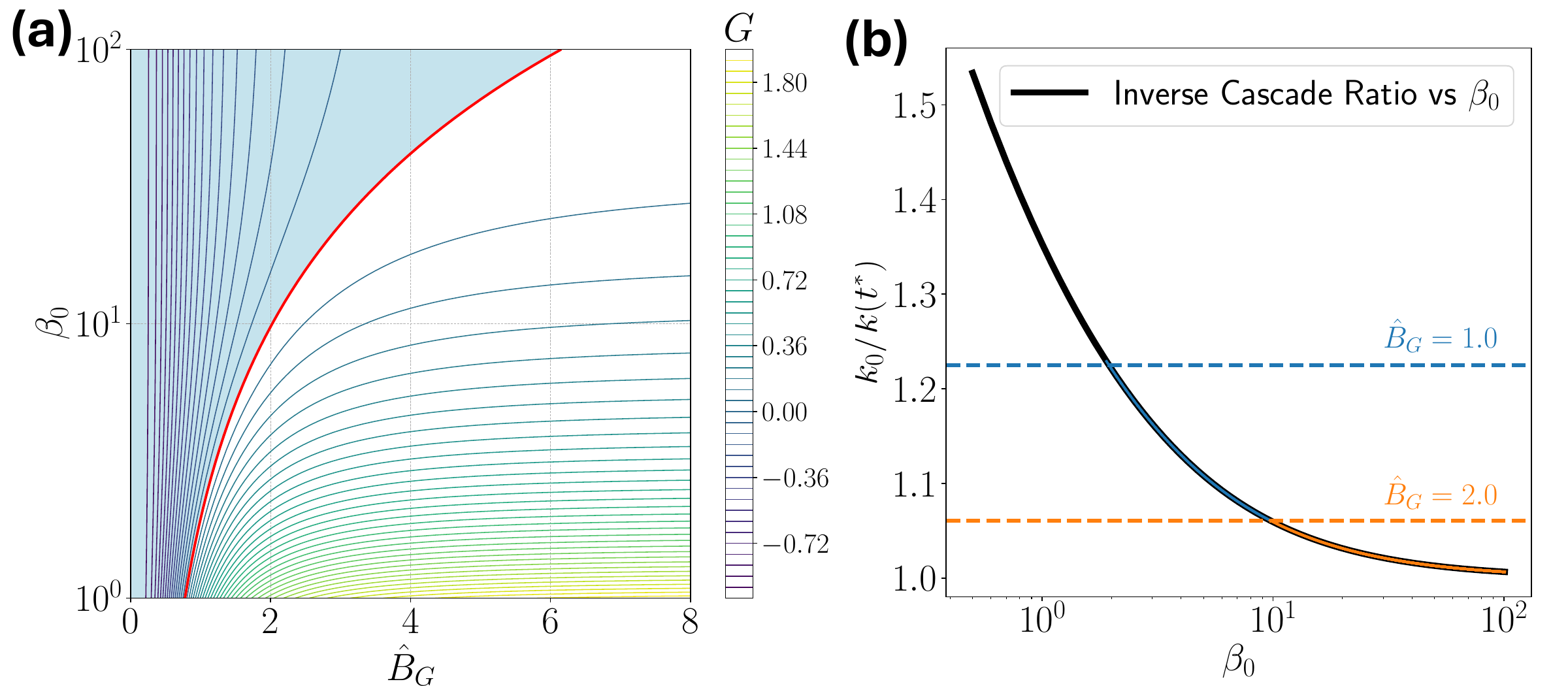}
\caption{\label{app:fig:4.Gcontour}(a) Contour plot of $G(\hat{B}_G, \beta_0)$. The red curve denotes the locus where $G(\hat{B}_G, \beta_0) = 0$, and the shaded region marks the parameter space in which inverse transfer is suppressed by the firehose instability. 
(b) Inverse cascade ratio as a function of plasma beta if firehose condition is triggered, obtained by solving Eq.~\eqref{app:eq:tstar}. Only the portions of the curve lying below the horizontal lines associated with a given guide field $\hat{B}_G$ represent valid solutions (see Eq.~\eqref{app:eq:valid}). }
\end{figure}

\begin{figure}[!htbp]
\centering
\includegraphics[width=1.0\textwidth]{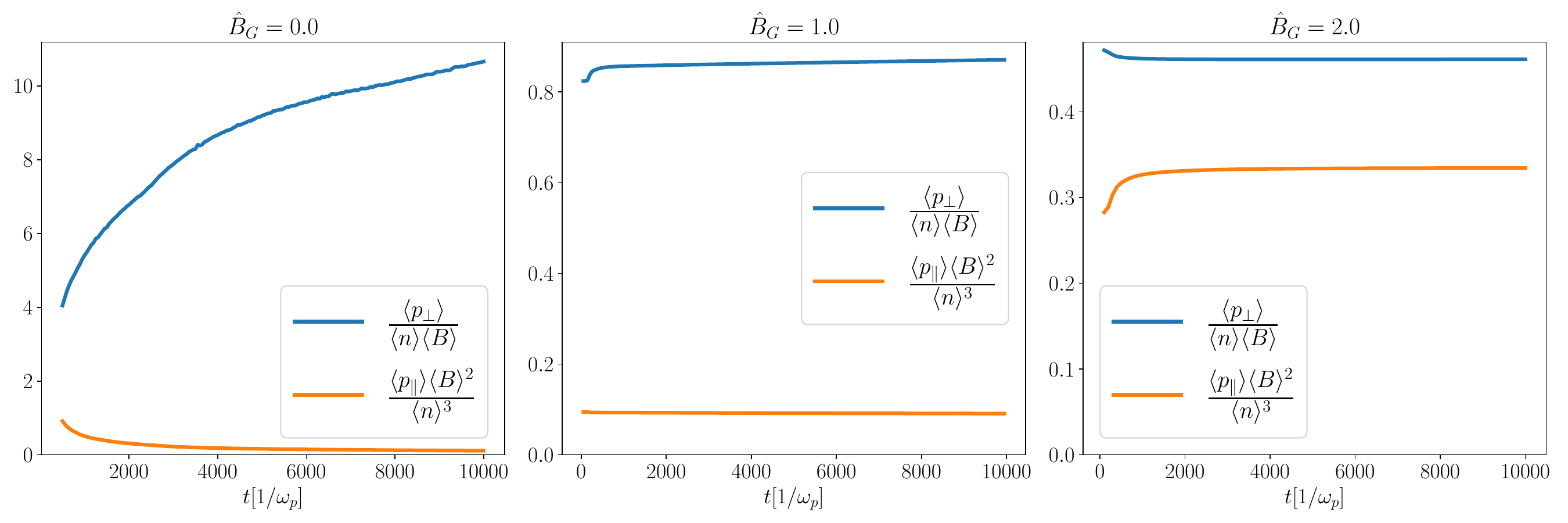}
    \caption{\label{app:fig:9.Atrace}  Time evolution of the two system-averaged CGL invariants for the three simulations discussed in the main text. For clarity, the orange curve of the $\hat{B}_G=0.0$ simulation has been multiplied by a factor of 100.}
\end{figure}

A contour plot of $G(\hat{B}_G,\beta_0)$ is shown in Fig.~\ref{app:fig:4.Gcontour} (a).
The shaded area marks the region of parameter space that is predicted to lead to the triggering of the firehose instability. 
It can be seen that in order to avoid firehose activation at larger values of ambient plasma beta requires progressively larger values of guide field.

If the firehose condition is eventually be met, the time at which the system will hit the firehose boundary, $t^*$, can be calculated from
\begin{equation}
\label{app:eq:tstar}
 \langle \Delta (t^*) \rangle +\frac{2}{\langle \beta_{\|}(t^*) \rangle}  \approx \left( \frac{\langle B(t^*) \rangle}{\langle B(t_0) \rangle}\right)^3 +\frac{2}{\beta_0} \left( \frac{\langle B(t^*) \rangle}{\langle B(t_0) \rangle}\right)^4 - 1 = 0.
\end{equation}
The inverse cascade ratio $k_0/k (t^*)$, approximated by $ \langle B(t_0) \rangle/ \langle B(t^*) \rangle$, can thus be obtained by solving Eq.~\eqref{app:eq:tstar}; the solution being valid only if
\begin{equation}
\label{app:eq:valid}
    \frac{k_0}{k(t^*)} \approx \frac{\langle B(t_0) \rangle }{\langle B(t^*) \rangle} = \left(\frac{2 \hat{B}_G^2 + 1}{2\hat{B}_G^2+\left( 1+  \left(\frac{t^*-t_0 }{\tau_0} \right)\right)^{-\alpha}}\right)^{1/2} < \left(\frac{2 \hat{B}_G^2 + 1}{2\hat{B}_G^2}\right)^{1/2},
\end{equation}
or the firehose condition will not be met. This implies, unsurprisingly, that a stronger guide field helps avoid the firehose condition.
Fig.~\ref{app:fig:4.Gcontour}(b) shows the inverse growth ratio as a function of (initial) plasma beta $\beta_0$.
It can be seen that the inverse cascade ratio is small for a large range of plasma beta if the firehose condition will eventually triggered. 
This figure also illustrates that with a stronger guide field, fewer solutions remain valid, allowing a broader range of $\beta_0$ values to avoid firehose instability.

Fig.~\ref{app:fig:9.Atrace} shows the time evolution of the two system-averaged CGL invariants, 
$\langle p_{\perp}\rangle/(\langle n\rangle \langle B\rangle)$ and 
$\langle p_{\parallel}\rangle \langle B\rangle^{2}/\langle n\rangle^{3}$. 
In simulations with a finite guide field, these quantities are indeed (approximately) conserved, whereas in the run without a guide field they are not.
An early-time transient is evident, most prominently for $\hat{B}_G=2.0$.
This is because the initial configuration is only an exact (MHD) equilibrium in the absence of a guide field. The inclusion of the latter drives the system away from equilibrium, more strongly as the magnitude of the guide field increases.
During this adjustment in the $\hat{B}_G=2.0$ run, a transient rise in 
$\langle p_{\parallel}\rangle \langle B\rangle^{2}/\langle n\rangle^{3}$ and dropping  $\langle p_{\perp}\rangle/(\langle n\rangle \langle B\rangle)$ drive 
$\langle T_{\parallel}\rangle$ upward and $\langle T_{\perp}\rangle$ downward more strongly than predicted by simple theory. 
This explains the more negative anisotropy and larger $\beta_{\parallel}$ than predicted seen in Fig.~\ref{fig:3.contour}(i).
Crucially, once past this initial adjustment, both invariants are well conserved throughout the main evolution, and the transient does not alter our main conclusions.

\section{Well magnetized scenario with a guide field\label{app:sec:2}}

\begin{figure}[!htbp]
\centering
\includegraphics[width=0.6\textwidth]{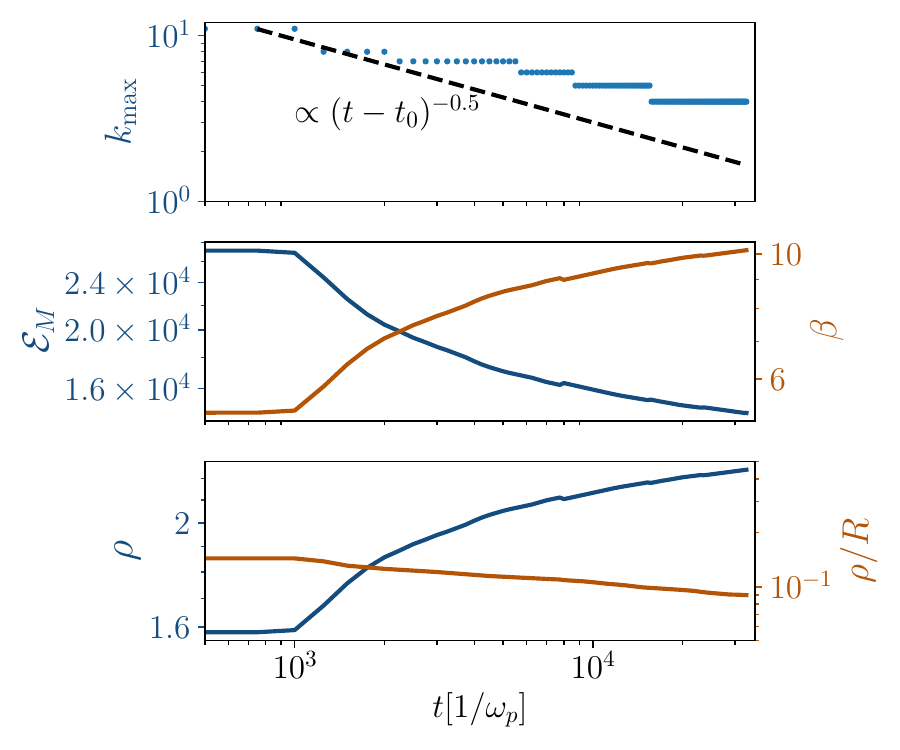}
\caption{\label{app:fig:8.trace_mhd_G} Same as Fig.~\ref{app:fig:5.trace_mhd} but with a guide field $\hat{B}_G = 1.0$.}
\end{figure} 

As discussed in the main text, the ``frozen'' scenario occurs when the magnetic structure size remains comparable to the scale at which firehose instability operates. 
As a result, magnetic tension, responsible for driving reconnection and island mergers, is effectively nullified. 
The resulting suppression of reconnection dynamics leads to the formation of elongated, worm-like magnetic structures and a halt in the inverse cascade. 

\begin{figure}[!htbp]
\centering
\includegraphics[width=0.8\textwidth]{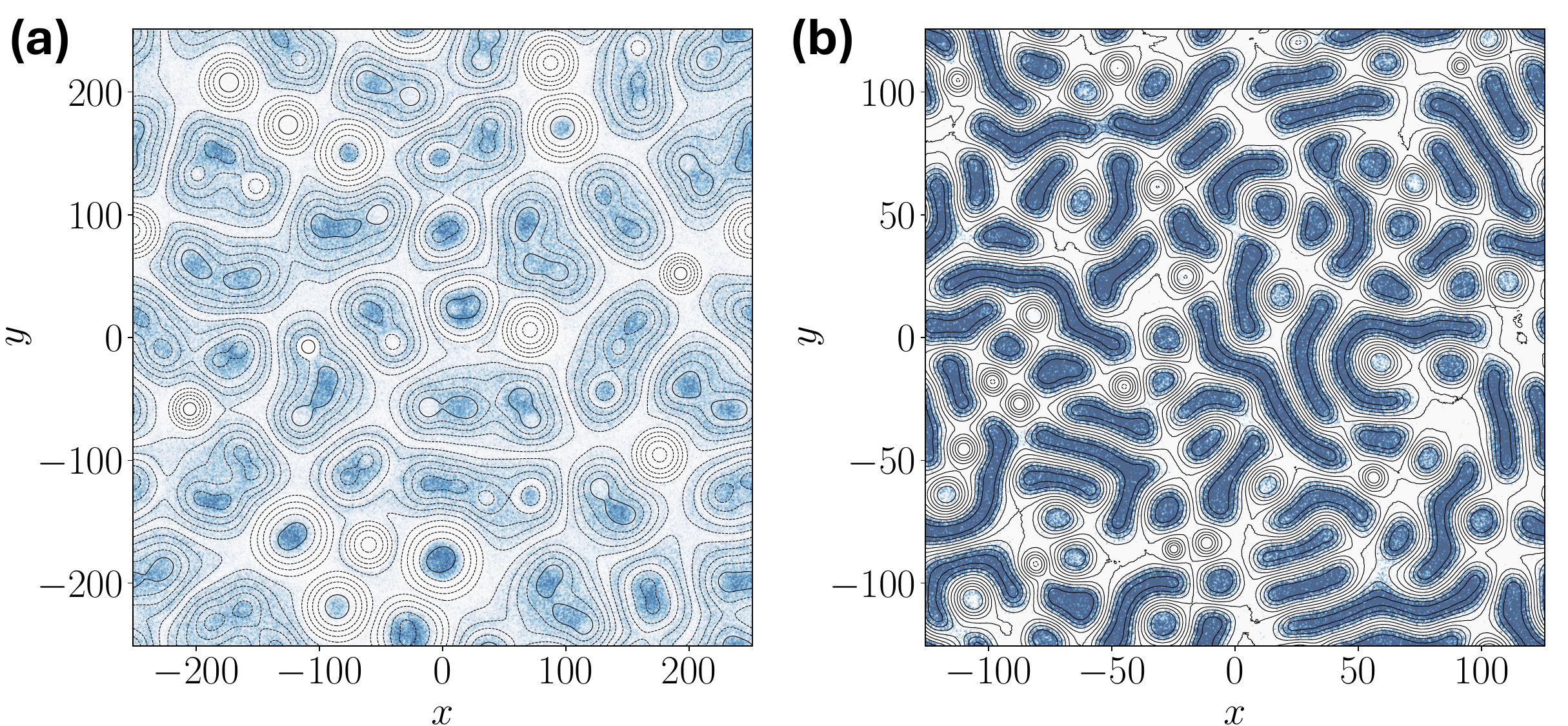}
\caption{\label{app:fig:6.unstable_contour} Contours of magnetic flux at $\omega_p t =10000$ for (a) well magnetized simulation and (b) marginally magnetized simulation (this simulation corresponds to the middle column of Fig.~\ref{fig:3.contour} as described in the main text).  The guide magnetic field strength is $\hat{B}_G=1.0$. Blue color marks the locations unstable to firehose instability. The color is binary (blue for unstable and white for stable locations); panel (b) appears darker because there are many more locations unstable to the firehose instability. }
\end{figure}

\begin{figure}[!htbp]
\centering
\includegraphics[width=0.5\textwidth]{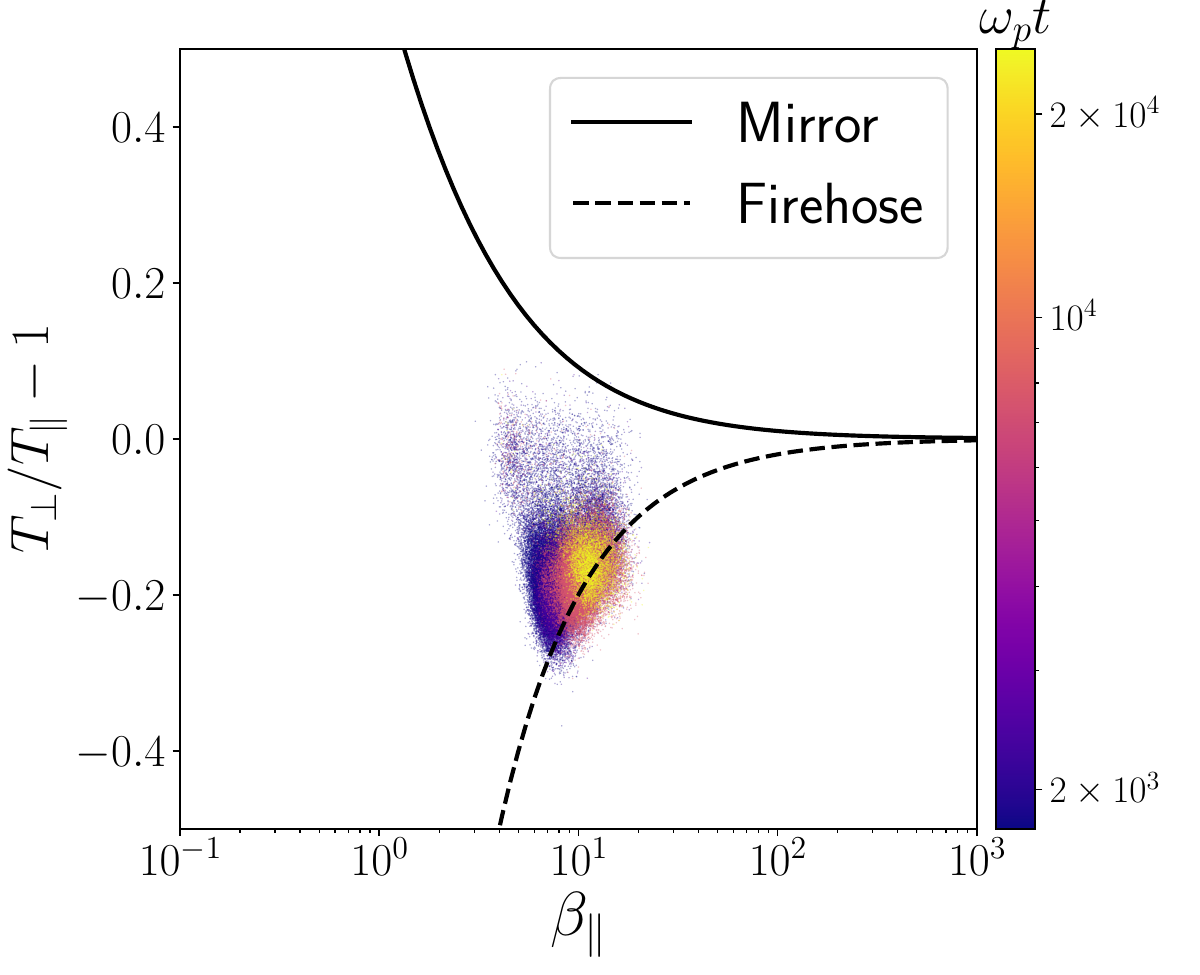}
\caption{\label{app:fig:7.ani_distribution} Scatter plots of pressure anisotropy versus parallel plasma beta at progressively later times for a well-magnetized plasma show that the system does not halt when it reaches the firehose threshold. Instead, it continues to evolve slowly along the firehose marginal stability boundary, moving gradually to higher $\beta_\parallel$ values. This indicates that magnetic mergers persist, albeit at a reduced rate, even in the presence of ongoing firehose activity. }
\end{figure}


To further confirm that this arrest only happens in marginally magnetized plasmas, we repeat the simulation in Appendix ~\ref{app:sec:3} with identical parameters but add a uniform guide field of magnitude $\hat{B}_G = 1.0$ (so that this simulation is identical to the simulation with $\hat{B}_G = 1.0$ in the main paper other than larger scale separation between $\rho$ and $R$).
Fig.~\ref{app:fig:8.trace_mhd_G} illustrates the time evolution of the key quantities in the simulation. Although the inverse cascade slows considerably at later times due to the onset of the firehose condition, it does not come to a complete halt.

Fig.~\ref{app:fig:6.unstable_contour} compares snapshots of the magnetic flux contours at $\omega_p t = 10000$ from the two simulations. 
Panel (a) corresponds to the well-magnetized case with the guide field (this simulation), and panel (b) shows the marginally magnetized case discussed in the main paper (the same snapshot as Fig.~\ref{fig:3.contour} (b)). 
The blue scatters in each panel mark locations that have crossed the firehose threshold, where the pressure anisotropy satisfies the instability condition. 
The color is binary (blue for unstable and white for stable locations).
In the marginally magnetized case, nearly all the elongated structures are firehose-unstable, highlighting how mergers concentrate the pressure anisotropy in precisely the regions where reconnection should proceed. 
In contrast, the well-magnetized simulation shows significantly fewer unstable locations, consistent with the notion that the stronger guide field suppresses firehose excitation in merger sites.

In addition to fewer regions being unstable to firehose instability in the well magnetized case, the instability saturates at the Larmor scale (much smaller than island size) and allows the plasma to relax into a marginally stable state \citep{kunz2014firehose}. 
As a result, magnetic tension is not eliminated at the island (reconnecting) scale.
This can also be observed in Fig.~\ref{app:fig:6.unstable_contour}. In the well-magnetized case, magnetic structures remain compact and round, indicative of sustained magnetic tension and active reconnection. 
In contrast, the marginally magnetized case exhibits strongly elongated structures with filamentary (``worm-like'') morphologies, consistent with nullified magnetic tension.
Consequently, reconnection and island coalescence continue when the plasma is well magnetized, though at a reduced rate. 
We observe that the long-term evolution proceeds along the firehose stability boundary, with the system gradually shifting to higher $\beta_\parallel$ values in parameter space, as shown in Fig.~\ref{app:fig:7.ani_distribution}.

These results demonstrate that while pressure-anisotropy–driven microinstabilities can completely arrest inverse cascades in marginally magnetized plasmas, the presence of a sufficiently strong magnetic guide-field --- and, therefore, the stronger magnetization of the plasma and consequent greater scale separation between the Larmor radius and the island size --- partially restores magnetic tension and permits continued reconnection and coalescence.

\bibliography{main}{}
\bibliographystyle{aasjournalv7}

\end{document}